\documentclass[a4paper,11pt]{article}
\pdfoutput=1 

\usepackage{jheppub} 

\usepackage[T1]{fontenc} 

\usepackage[utf8]{inputenc}
\usepackage{lmodern}

\usepackage{enumerate}
\usepackage{float}
\usepackage{color}
\usepackage{slashed}
\usepackage{multirow}

\usepackage{chngcntr}
\usepackage{amssymb, amsmath,mathrsfs}

\usepackage{graphics}
\usepackage{graphicx}
\usepackage{epsf}
\usepackage{epsfig}
\usepackage{float}
\usepackage{makecell}
\usepackage{multirow}
\usepackage{color}
\usepackage{xcolor}
\usepackage{subcaption}
\usepackage{tikz-feynman}

\usepackage[utf8]{inputenc}
\usepackage{amsmath}
\usepackage{amssymb}
\usepackage[normalem]{ulem}

\usepackage{mathtools}
\usepackage{amsmath}
\usepackage{adjustbox}

\newcommand\undermat[2]{
  \makebox[0.5pt][l]{$\smash{\underbrace{\phantom{%
    \begin{matrix}#2\end{matrix}}}_{ \let\scriptstyle\textstyle\text{\large $#1$}}}$}#2}
\newcommand\overmat[2]{
  \makebox[-1pt][l]{$\smash{\overbrace{\phantom{%
    \begin{matrix}#2\end{matrix}}}^{ \let\scriptstyle\textstyle\text{\large $#1$}}}$}#2}    
\usepackage{natbib}
\usepackage{tikz}
\usepackage[vcentermath]{youngtab}
\usepackage{slashed} 
\usepackage[font=small]{caption} 
\usepackage{times} 

\usepackage{bm}       
\usepackage{bbm} 

\usepackage{relsize}  

\usepackage{autobreak}  

\usepackage{makeidx} 
\usepackage{bbding} 
\usepackage{listings} 
\usepackage{ytableau}

\newcolumntype{M}[1]{>{\centering\arraybackslash}m{#1}}
\newcolumntype{N}{@{}m{0pt}@{}}

\newcommand{\comments}[1]{}
\newcommand{\ab}[1]{\langle #1   \rangle}

\newcommand{\beq}{\begin {equation}}  
\newcommand{\eeq}{\end   {equation}} 
\newcommand{\bea}{\begin {eqnarray}} 
\newcommand{\eea}{\end   {eqnarray}}  
\newcommand{\baa}{\begin {array}   } 
\newcommand{\eaa}{\end   {array}   }     
\newcommand{\bit}{\begin {itemize} }
\newcommand{\eit}{\end   {itemize} }
\newcommand{\be }{\begin {equation}} 
\newcommand{\ee }{\end   {equation}}

\usepackage{mmacells}

\mmaDefineMathReplacement[≤]{<=}{\leq}
\mmaDefineMathReplacement[≥]{>=}{\geq}
\mmaDefineMathReplacement[≠]{!=}{\neq}
\mmaDefineMathReplacement[→]{->}{\to}[2]
\mmaDefineMathReplacement[⧴]{:>}{:\hspace{-.2em}\to}[2]
\mmaDefineMathReplacement{∈}{\in}
\mmaDefineMathReplacement{∉}{\notin}
\mmaDefineMathReplacement{∞}{\infty}
\mmaDefineMathReplacement{†}{\dagger}
\mmaDefineMathReplacement{ψ}{\psi}
\mmaDefineMathReplacement{⊗}{\otimes}
\mmaDefineMathReplacement{τ}{\tau}
\mmaDefineMathReplacement{λ}{\lambda}
\mmaDefineMathReplacement{Γ}{\Gamma}

\mmaSet{
  morefv={gobble=2},
  linklocaluri=mma/symbol/definition:#1,
  morecellgraphics={yoffset=1.9ex}
}

\title{\boldmath Partial-Wave Unitarity Bounds on Higher-Dimensional Operators from 2-to-$N$ Scattering}

\author[b]{C\'eline Degrande,}
\author[a,b]{Hao-Lin Li,}
\author[c]{Ling-Xiao Xu.}

\affiliation[a]{School of Physics, Sun Yat-Sen University, Guangzhou 510275, P. R. China}
\affiliation[b]{Centre for Cosmology, Particle Physics and Phenomenology (CP3), Universite Catholique de Louvain}
\affiliation[c]{Abdus Salam International Centre for Theoretical Physics, Strada Costiera 11, 34151, Trieste, Italy}

\emailAdd{lihlin68@mail.sysu.edu.cn}
\emailAdd{celine.degrande@uclouvain.be}
\emailAdd{lxu@ictp.it}

\abstract{We present a systematic method for deriving partial-wave unitarity bounds on Wilson coefficients of higher-dimensional operators in effective field theories involving more than four fields, which naturally appear in tree-level 2-to-$N$ scattering processes with $N \geq 3$. Unlike 2-to-2 scattering, 2-to-$N$ scattering with $N \geq 3$ features multiple amplitudes associated with the same total angular momentum. To resolve these degeneracies, we provide a way to construct an orthonormal amplitude basis by parameterizing the phase space manifold of massless particles using spinor-helicity variables, enabling analytical integration over the phase space with arbitrary particle numbers. We provide Mathematica code to analytically evaluate phase space integrals of interference between two local on-shell amplitudes up to four final-state particles, with straightforward generalization to $N$ final-state particles. As practical applications, we demonstrate the use of this tool by deriving unitarity bounds on some dimension-7 and dimension-8 operators in the Standard Model effective field theory involving five and six fields, respectively.
}

\begin{document} 
\maketitle
\flushbottom

\section{Introduction}
In the post-Higgs-discovery era, the Standard Model effective field theory (SMEFT), due to its universality and systematic nature, has become increasingly popular as a theoretical framework for analyzing small deviations from Standard Model (SM) predictions caused by hypothetical ultraviolet (UV) theories at high energy scales.

Recently, rapid developments on the theoretical side --- such as the automation of operator basis construction, matching computation, and global fit analysis --- have paved the way for systematic phenomenological studies involving a large number of higher-dimensional operators beyond the dimension-six level. (For a recent review on computation tools for effective field theories, see~\cite{Aebischer:2023nnv}). 
It has also been shown that several dimension-8 SMEFT operators can be probed at future collider experiments~\cite{Degrande:2013kka,Bellazzini:2018paj,Ellis:2021dfa,Ellis:2022zdw,Boughezal:2022nof,Boughezal:2023nhe,Ellis:2023ucy,Ellis:2023zim,ElFaham:2024egs,Degrande:2024mbg,Martin:2023tvi,Assi:2024zap,Gillies:2024mqp,Ellis:2025ghl}. 
However, to interpret the experimental results as the constraints or measured values of Wilson coefficients, it is essential to ensure that the underlying theoretical predictions within the SMEFT framework are consistent. 
Therefore, theoretical consistency conditions --- such as partial-wave unitarity bounds and positivity bounds~\cite{deRham:2022hpx}, which are derived from unitarity and locality of the S-matrix --- impose important constraints on the allowed structure and size of EFT operators. 
Indeed, for a Bayesian global fit, these theoretical constraints can also serve as priors on the Wilson coefficients from the beginning of the statistical analysis, especially if one aims to filter out the UV theories that violate fundamental principles.

Positivity bounds in the SMEFT have been extensively studied in recent years~\cite{Zhang:2018shp, Bellazzini:2018paj, Remmen:2019cyz, Remmen:2020vts, Bonnefoy:2020yee, Fuks:2020ujk, Gu:2020ldn, Davighi:2021osh, Li:2022rag, Ghosh:2022qqq, Chen:2023bhu, Gu:2023emi, Chala:2023xjy}. 
For instance, it was shown in~\cite{Zhang:2018shp} that applying positivity constraints to the parameter space of dimension-8 operators relevant to anomalous quartic gauge couplings excludes approximately 98\% of the experimentally allowed region.
However, such positivity bounds can only restrict the Wilson coefficients to a convex hull in the parameter space, and typically they apply to the 2-to-2 scattering amplitudes.
In contrast, partial-wave unitarity bounds play a complementary role by providing the upper limits on the Wilson coefficients~\cite{Chen:2023bhu}.
Perturbative partial-wave unitarity bounds for dimension-6 SMEFT operators have been studied extensively in e.g.~\cite{Baur:1987mt, Gounaris:1995ed, Layssac:1993vfp, Giudice:2007fh, Corbett:2014ora, Corbett:2017qgl}, all focusing on the 2-to-2 scattering processes. 
However, for operators of dimension higher than six, many of them contain more than four fields, and their leading contributions naturally appear in scattering amplitudes involving more than four external particles. In such cases, it becomes necessary to extend the analysis of partial-wave unitarity to general 2-to-$N$ scattering processes.
For instance, an upper bound on the scale responsible for neutrino mass generation can be derived from the scaling of the cross sections of 2-to-$N$ processes~\cite{Maltoni:2001dc}, while the analysis only relies on the $J=0$ partial wave.
Similar result is obtained from the scattering of one neutrino and $N$ Goldstone to themselves in Ref.~\cite{Ekhterachian:2021rkx}, by generalizing the unitatiry bound  for the scattering of $N$ scalars of different species in~\cite{Chang:2019vez}.
Furthermore, the $J=0$ unitarity bounds on the EFT operators from the scattering process $VV\to hhh$ were also studied~\cite{Kilian:2018bhs}. Until recent years, the partial-wave amplitude basis was constructed~\cite{Jiang:2020rwz}, and the partial-wave unitarity bounds for SMEFT operators beyond dimension six were studied using 2-to-3 scattering processes~\cite{Bresciani:2025toe}, where a list of 2-to-3 partial-wave helicity amplitudes (of the lowest mass dimension) was also provided explicitly. Recently, the authors of ~\cite{Bresciani:2025toe} also derived the unitarily bound and compared with the positivity point in axion-like particle EFT up to dimension-8~\cite{Bresciani:2025ojh}.

More technically, the study of generalized 2-to-$N$ partial-wave amplitudes has been relatively limited, mainly due to the following two challenges. First, unlike the 2-to-2 case, where the partial-wave basis of massless particles is constructed using the well-known Wigner $d$-functions, the general local on-shell amplitude basis for 2-to-$N$ processes was not known until recently. This challenge was addressed in~\cite{Jiang:2020rwz}, where a method based on the action of the Casimir operator of the Poincaré group was introduced. Second, for general 2-to-$N$ scattering, the partial-wave basis can be degenerate for a given total angular momentum $J$. Consequently, obtaining an orthogonal set of basis EFT amplitudes and determining their normalization requires performing the full $N$-body phase space integration analytically. 

For the massless $N$-body phase space, Refs.~\cite{Zwiebel:2011bx,EliasMiro:2020tdv} provide an explicit parameterization for the three-body final state by expressing the spinor-helicity variables of the final-state particles as complex linear combinations of those of the two initial-state particles. Using a similar approach, Ref.~\cite{Larkoski:2020thc} nearly achieves a complete parameterization of the massless $N$-body phase space that resolves all delta functions arising from momentum conservation. However, an explicit expression for the transformation matrix between the two sets of integration variables at an intermediate step is missing.
Ref.~\cite{Cox:2018wce} also proposes an $N$-body phase space parameterization, but it is primarily designed for Monte Carlo phase-space generation and the treatment of infrared divergences, making it less convenient for the analytical computations considered in this work. Consequently, a fully explicit and convenient parameterization of the general $N$-body massless phase-space integral, suitable for analytical computation for the local on-shell amplitudes, has not yet been presented in the literature.

In this paper, we begin by reviewing the basic concepts behind the Casimir operator method, which enables the systematic construction of the massless partial-wave amplitude basis. This corresponds to the bipartite case of the general J-basis amplitudes. Through an explicit example, we demonstrate how these amplitudes can be derived using a single line of code with the Mathematica package \texttt{ABC4EFT}~\cite{Li:2022tec}.

Next, we highlight a key observation: the phase space integral of an $N$-body massless final state can be computed analytically using a parametrization inspired by spinor-helicity variables in Ref.~\cite{Larkoski:2020thc}.
We find that the missing piece of the explicit transformation matrix in that paper can be analytically obtained by Sherman-Morrison formula.
This enables us to write phase space integration in closed analytic form using a set of spherical coordinates.
To our knowledge, this is the first derivation of an explicit analytic parametrization and phase space measure for a 4-body massless final state. 
Moreover, our method generalizes straightforwardly to the case of $N$-body final states. 
Based on this parameterization, we automate the evaluation of phase space integrals for the interference of local on-shell amplitudes in 2-to-3 and 2-to-4 scattering processes. We provide the Mathematica code to assist the reader in performing these integrals analytically.
Finally, using these tools, we demonstrate --- through explicit examples --- how to derive partial-wave unitarity bounds for SMEFT dimension-7 and dimension-8 operators with five and six fields, respectively. 

This paper is organized as follows. In section.~\ref{sec:2}, we derive the master formula of unitarity constraints for the generalized 2-to-$N$ amplitudes starting from the 2-to-2 scattering process. In section.~\ref{sec:partial-wave}, we review the formalism for constructing the partial-wave amplitudes. In section.~\ref{sec:phase-space}, we introduce the general procedure for constructing the orthonormal partial-wave amplitude basis for a $N$-body final state, and we illustrate our general procedure with concrete examples of 3-body and 4-body cases. In section.~\ref{sec:smeft_example}, we discuss how to derive the partial-wave unitarity bounds on the Wilson coefficients for the SMEFT operators with more than four fields. Finally, we conclude in section.~\ref{sec:conc}. In appendix.~\ref{sec:appA} we summarize our conventions for spinor variables. In appendix.~\ref{sec:appB}, we show how to parameterize the massless $N$-body phase space using a set of spherical coordinates, which enables us to perform the integral analytically.

\section{Partial-wave unitarity bounds from $2\to N$ scattering}\label{sec:2}

The scattering matrix is defined as $S=\mathbbm{1}+i{\cal T}$, where ${\cal T}$ is related to the scattering amplitude $M_{i\to f}$ as
\begin{eqnarray}
    \langle f|T|i\rangle = (2\pi)^4\delta^4(p_f-p_i)M_{i\to f},
\end{eqnarray}
where $i$ and $f$ denote the initial and final states, respectively, specified by the four-momenta, helicities, and all internal quantum numbers of the particles involved in the scattering process.  
The unitarity condition $S^\dagger S=\mathbbm{1}$ implies that ${\cal T}-{\cal T}^\dagger = i{\cal T}^\dagger {\cal T}$. As a result, the scattering amplitude satisfies the condition
\begin{eqnarray}
    M_{i\to f}-M_{f\to i}^* = i\sum_{X}\int d\Pi_X M^*_{f\to X}M_{i\to X}(2\pi)^4\delta^{4}(p_i-p_X),\label{eq:unitarity1}
\end{eqnarray}
where we used the identity $\langle f| {\cal T}^\dagger |i \rangle = \langle i |{\cal T}|f \rangle ^*=M^*_{f\to i}(2\pi)^4\delta^4(p_i-p_f)$. The phase space measure for the intermediate particles is defined as $d\Pi_X = \prod_j\frac{d^3p_j}{(2\pi)^32E_j}$.
Throughout the paper, we consider only the scattering amplitudes of massless particles. In the context of SMEFT, this corresponds to the unbroken phase of electroweak symmetry. We assume that taking the massless limit is a valid approximation for studying the energy scale at which perturbative unitarity is violated. Indeed, due to the decoupling nature of heavy particles, unitarity violation in SMEFT is expected to occur only at sufficiently high energies.

To derive the general partial-wave unitarity bound for $2$-to-$N$ scattering, we begin by considering the case where both the initial and final states, $i$ and $f$, are two-particle states. In this case, the scattering amplitude is fully determined by the center-of-mass energy $s$ and the scattering angle $\theta$. 
For forward elastic scattering (i.e. $\theta=0$), eq.~\eqref{eq:unitarity1} becomes
\begin{eqnarray}\label{eq:zeroscattering}
    2 {\rm Im}M(0) = \sum_{X\neq i}\int d\Pi_X|M_{i\to X}|^2(2\pi)^4\delta^4(p_X-p_i) + \int \frac{\sin\theta d\theta}{16\pi} |M(\theta)|^2,
\end{eqnarray}
where we have separated the contributions from the intermediate states $X\neq i$ from those where $X=i$, they correspond to the inelastic scattering and the elastic ones, respectively. Here, $M(\theta)$ denotes the elastic scattering amplitude for a specific 2-to-2 process involving massless particles. 

For a specific scattering process with a definite helicity configuration such as $A_{\lambda_1}B_{\lambda_2}\to A_{\lambda_1}B_{\lambda_2}$, the amplitude $M(\theta)$ can be expanded in particle waves with definite total angular momentum $J$ as
\begin{eqnarray}
    &&M(\theta) = 16\pi\sum_J(J+1/2) d^{J}_{\lambda\lambda}(\theta) T^J(s),
\end{eqnarray}
where $d^{J}_{\lambda\lambda}$ is the Wigner $d$-matrix and $\lambda=\lambda_1-\lambda_2$ is the helicity difference in the initial state. 
Substitute the partial-wave expansion of $M(\theta)$ into the second term on the right-hand side of eq.~\eqref{eq:zeroscattering} yields
\begin{eqnarray}
    \int \frac{\sin\theta d\theta}{16\pi} |M(\theta)|^2 &=\sum_J 16\pi(J+1/2)|T^J(s)|^2,
\end{eqnarray}
where we have used the orthonormality condition for the Wigner $d$-matrix:
\begin{eqnarray}
    \int d_{\lambda\mu}^Jd_{\lambda\mu}^{J'}d\cos\theta = \frac{\delta^{JJ'}}{(J+1/2)}.
\end{eqnarray}
Likewise, the left-hand side of eq.~\eqref{eq:zeroscattering} can be decomposed into partial waves as
\begin{eqnarray}
    2\ {\rm Im} M(0) =2\sum_J 16\pi (J+1/2)\ d^J_{\lambda\lambda}(0) \ {\rm Im} T^J(s) \;.
\end{eqnarray}
Using $ d^J_{\lambda\lambda}(0)=1$, 
eq.~\eqref{eq:zeroscattering} becomes
\begin{eqnarray}
    2\sum_J16\pi(J+1/2){\rm Im}T^J(s) = \sum_{X\neq i} \int d\Pi_X|M_{i\to X}|^2(2\pi)^4\delta^4(p_X-p_i)+\sum_J 16\pi (J+1/2)|T^J(s)|^2. \nonumber\\
    \label{eq:unitarity2}
\end{eqnarray}
Since the S-matrix is block-diagonal in total angular momentum $J$ (i.e., angular momentum conservation), one can isolate each $J$-block in eq.~\eqref{eq:unitarity2}. 
For a 2-to-2 scattering process with definite total angular momentum $J$, we have the partial-wave unitarity bound as
\begin{eqnarray}
    |T^J|\leq 2,\quad 0\leq {\rm Im}T^J\leq 2,\quad |{\rm Re} T^J|\leq 1.
\end{eqnarray}
This result is well known in the literature.

We now derive the partial-wave unitarity bounds for $2$-to-$N$ scattering from eq.~\eqref{eq:unitarity2}, a result much less explored in the literature. Let us begin by decomposing the amplitude as 
\begin{equation}
M_{i\to X} = \sum_{J,a}C^{Ja}_{i\to X} B^{Ja}_{i\to X}\;, 
\end{equation}
where $J$ denotes the total angular momentum with the corresponding partial-wave amplitude $B^{Ja}_{i\to X}$, which is a function of the four-momenta of all the external particles. As we will see in concrete examples in section~\ref{sec:partial-wave}, the partial-wave amplitudes for a fixed $J$ may not be unique. Consequently, we introduce an additional index $a$ to label the degeneracy for the partial-wave amplitudes with fixed $J$. All other relevant information, such as the Wilson coefficients and gauge quantum numbers, is included in $C^{Ja}_{i\to X}$.
We assume that the basis amplitudes $B^{Ja}_{i\to X}$ satisfy the following orthonormality condition
\begin{eqnarray}
    \int d\Pi_X B^{Ja}_{i\to X}(B^{J'a'}_{i\to X})^*(2\pi)^4\delta^4(p_X-p_i) \equiv  \Big\langle B^{J'a'}_{i\to X}, B^{Ja}_{i\to X}\Big\rangle= g^{Ja}_{i\to X}(s)\delta_{aa'}\delta_{JJ'},\label{eq:unitarity3}
\end{eqnarray}
it means that amplitudes with different quantum numbers $J$ and $a$ are orthogonal to each other after integrating over the $N$-body phase space of the intermediate state.
It follows that, for each partial wave $J$, the unitarity condition in eq.~\eqref{eq:unitarity2} becomes
\begin{eqnarray}
    2{\rm Im}T^J(s) = \sum_{X\neq i}\frac{\sum_a g_{i\to X}^{Ja}(s)|C_{i\to X}^{Ja}|^2}{16\pi(J+1/2)} + |T^J(s)|^2.
    \label{eq:unitarity2p}
\end{eqnarray}
From eq.~\eqref{eq:unitarity2p}, we obtain the master formula for the partial-wave unitarity bounds for $2$-to-$N$ scattering amplitudes:
\begin{eqnarray}
    \frac{\sum_{a,X\neq  i } g_{i\to X }^{Ja}(s)|C_{i\to X}^{Ja}|^2}{16\pi(J+1/2)}\leq 1.\label{eq:unitarity4}
\end{eqnarray}

\section{Constructing partial-wave amplitudes}
\label{sec:partial-wave}

As seen from eqs.~\eqref{eq:unitarity3} and~\eqref{eq:unitarity4}, once the basis for the partial-wave amplitudes $B^{Ja}_{i\to X}$ and their corresponding normalization factors $g^{Ja}_{i\to X}(s)$ are determined, one can derive the partial-wave unitarity bounds on the Wilson coefficients that appear in the amplitude for the process $i\to X$. 
Therefore, the first task is to systematically construct the basis of $B^{Ja}_{i\to X}$, which can be achieved by taking the following steps:
\begin{enumerate}
\item For a given mass dimension, we first systematically construct a basis of local, on-shell amplitudes with the Young Tensor Method~\cite{Henning:2019enq, Li:2020gnx} using spinor-helicity variables. The amplitude basis obtained as such is denoted as Y-basis.
The advantage of this method is that it automatically removes the redundancies from momentum conservation and the Schouten identity.
\item Within the space of the complete Y-basis amplitudes, we determine the matrix representation of the Casimir operator $W^2=W_\mu W^\mu$ of the Poincar\'e group. 

Here, $W^\mu$ is the Pauli-Lubanski vector defined as $W_\mu = \frac{1}{2}\epsilon_{\mu\nu\rho\sigma}P^\mu M^{\rho\sigma}$, where $P^\mu$ is the momentum operator and $M^{\mu\nu}$ is the Lorentz group generator. 
In terms of spinor-helicity variables, we have
\begin{eqnarray}
     W^2 &=\frac{1}{8} P^2(\mathrm{Tr}[M^2] +\mathrm{Tr}[\widetilde{M}^2]) -\frac14 \mathrm{Tr}[P^\intercal M P\widetilde{M}] \, .
\end{eqnarray}
where $P=P_\mu \sigma^\mu_{\alpha\dot\alpha}$, $P^\intercal=P_\mu\bar\sigma^{\mu\dot\alpha\alpha}$, and $M$, $\widetilde{M}$ are the chiral components defined as $ M_{\mu\nu} \sigma^\mu_{\alpha\dot{\alpha}} \sigma^\nu_{\beta\dot{\beta}} =\epsilon_{\alpha\beta}\widetilde{M}_{\dot{\alpha}\dot{\beta}} +\tilde{\epsilon}_{\dot{\alpha}\dot{\beta}} M_{\alpha\beta}$.
For a given subset ${\cal I}$ of particles in a scattering process, $\widetilde{M}_{\dot{\alpha}\dot{\beta}}$ and $M_{\alpha\beta}$ act on amplitudes as the differential operators:
\begin{eqnarray}
    M_{\alpha \beta}=&i \sum_{i\in {\cal I}}\left(\lambda_{i \alpha} \frac{\partial}{\partial \lambda_{i}^{\beta}}+\lambda_{i \beta} \frac{\partial}{\partial \lambda_{i}^{\alpha}}\right),\quad \widetilde{M}_{\dot{\alpha} \dot{\beta}}=i \sum_{i\in {\cal I}}\left(\tilde{\lambda}_{i \dot{\alpha}} \frac{\partial}{\partial \tilde{\lambda}_{i}^{\dot{\beta}}}+\tilde{\lambda}_{i \dot{\beta}} \frac{\partial}{\partial \tilde{\lambda}_{i}^{\dot{\alpha}}}\right).
\end{eqnarray}
It is the summation over the subset ${\cal I}$ of particles that enables us to define the specific scattering channel with definite angular momentum. 
\item We obtain the partial-wave amplitudes $B^{Ja}_{{\cal I}\to X}$ as the eigenvectors of the matrix representation of $W_{\cal I}^2$, such that $W_{\cal I}^2B^{Ja}_{{\cal I}\to X}=-s_{\cal I}J(J+1)B^{Ja}_{{\cal I}\to X}$, where $s_{\cal I}\equiv (\sum_{i\in {\cal I}}p_i)^2$. This set of partial-wave amplitudes forms the J-basis for the given scattering channel ${\cal I}$.  (In the rest of the paper, we will use the terms "partial-wave amplitudes" and "J-basis amplitudes" interchangeably.)
\end{enumerate}

For clarity, we illustrate the steps described above with a simple example. 
Let us consider the scattering process involving five massless particles $\psi_1\phi_2\phi_3\phi_4\psi_5^\dagger$, where $\phi$, $\psi$, $\psi^\dagger$ represent particles with helicities $0$, $-1/2$, and $1/2$, respectively. The subscripts of these fields correspond to the labeling of their momenta. 
\begin{enumerate}
\item Using the Young Tensor Method, we identify two amplitudes of mass dimension two in the Y-basis
\begin{eqnarray}
    B^y_1=\ab{14}[45],\quad B^{y}_2 = -[35]\ab{13}\;,
\end{eqnarray}
where all the particles are considered to be incoming.
\item Suppose we focus on partial-wave amplitudes for the particles $\phi_2\phi_3$. Acting the Casimir operator $W^2_{23}$ on the amplitudes $B^y_1$ and $B^{y}_2$ yields the following amplitudes of mass dimension four
\begin{eqnarray}
    W^2_{23} B_1^y &&= 0,\\
    W^2_{23} B_2^y &&= s_{23}(B_1^y-2B_2^y). 
\end{eqnarray}
Importantly, the right-handed side of these equations can be rewritten as linear combinations of the original amplitudes $B^y_1$ and $B^{y}_2$ after factoring out the kinematic variable $s_{23}=(p_2+p_3)^2$. Hence, the matrix representation of $({\cal W}^2_{23})_{ji}$,  defined by $ W^2_{23}B^y_i=B^y_j({\cal W}^2_{23})_{ji}$ on the basis spanned by $B^y_{1,2}$, is
\begin{eqnarray}
    {\cal W}^2_{23} = s_{23}\begin{pmatrix}
        0 & 1\\
        0&-2
    \end{pmatrix}.
\end{eqnarray}
\item Diagonalizing the above matrix ${\cal W}^2_{23}$ yields two eigenvalues, $-2s_{23}$ and $0$, and they correspond to the angular momentum channels $J=0$ and $J=1$, respectively. The associated eigenvectors are
\begin{eqnarray}
     B^{J=0}: &&\ab{14}[45] ,\nonumber \\
     B^{J=1}: &&-2\ab{13}[35] - \ab{14}[45].\label{eq:ampJ1}
 \end{eqnarray}
These eigenvectors $B^{J=0}$ and $B^{J=1}$ are the J-basis amplitudes for the particles $\phi_{2,3}$.
\end{enumerate}

Unlike 2-to-2 scattering, one prominent feature of the J-basis amplitudes for 2-to-$N$ processes with $N\geq 3$ is that, for a fixed total angular momentum, the corresponding J-basis amplitudes can be non-unique. 
We illustrate this point with the amplitudes of mass dimension 2 in the scattering process of five scalar particles, which corresponds to the effective operator of the type $\phi^5 D^2$.
To start, we have the Y-basis amplitudes
\begin{eqnarray}
    B^y_1=-s_{45},\ B^y_2=s_{35},\ B^y_3=-s_{34},\ B^y_4=-s_{25},\ B^y_5=s_{24}, 
\end{eqnarray}
where $s_{ij}=(p_i+p_j)^2$, and all the particles are the incoming ones. The corresponding J-basis amplitudes for the scattering channel initiated by $\phi_1\phi_2$ are given by
\begin{eqnarray}
    &&B^{J=0, a=1,2,3}_{0,0;0,0,0}: -s_{34},\ s_{35},\ -s_{45},\label{eq:jbphi51}\\
     &&B^{J=1, a=1,2}_{0,0;0,0,0}: 2s_{24}+s_{34}+s_{45},\ -2 s_{25}-s_{35}-s_{45},\label{eq:jbphi52}
\end{eqnarray}
where both the $J=0$ and $J=1$ partial-wave amplitudes are degenerate.
 
The procedure for obtaining the J-basis amplitudes described above is fully automated using the Mathematica package \texttt{ABC4EFT}~\cite{Li:2022tec}. To be specific, this is realized by the function 
\mmaInlineCell[pattern={k,ch,state},defined=W2Diagonalize]{Code}{W2Diagonalize[state,k,ch]}, which requires three inputs: "state", the particle configuration listed in an order of increasing helicities; "k", the number of derivatives in the corresponding operator; "ch", the subset of particles specifying the scattering channel for which partial-wave amplitudes of definite angular momentum are constructed.
For example, the J-basis amplitudes shown in eq.~\eqref{eq:ampJ1} can be obtained by the following command:
\begin{mmaCell}[defined={W2Diagonalize,Ampform}]{Code}
  W2Diagonalize[{-1/2, 0,0,0, 1/2}, 1, {2, 3}] //Ampform
\end{mmaCell}
\begin{mmaCell}[defined={ab,sb}]{Output}
  <|"basis"->\big\{[45]<14>,-[35]<13>\big\}, "j"->\{1, 0\}, \\"transfer"->\{\{-1, 2\}, \{1, 0\}\}, "j-basis"->\big\{-2[35]<13>-[45]<14>,\\[45]<14>\big\}|>
\end{mmaCell}
The output contains four keys: "basis", the complete and independent Y-basis of amplitudes constructed with the Young tensor method; "j", a list of total angular momentum values allowed for the specified scattering channel; "transfer", the transformation matrix from the Y-basis to the J-basis; "j-basis", the concrete J-basis amplitudes for the scattering channel, ordered according to the entries in the "j" list.
In this way, one can verify the partial-wave 5-point amplitudes (up to normalization) that appear in the appendix of~\cite{Bresciani:2025toe}, where only the local on-shell amplitudes of the lowest dimension are included for each helicity configuration.

\section{Phase space integral and normalization of partial-wave amplitudes}
\label{sec:phase-space}
In the previous section, we introduced a systematic method for constructing the partial-wave amplitude basis for scattering processes involving more than two particles in the final state. To obtain the partial-wave unitarity bound as given in eq.~\eqref{eq:unitarity4}, it remains to compute the normalization factor $g^{Ja}_{i\to X}(s)$, which requires performing the $N$-body phase space integral over the final-state massless particles, as shown in eq.~\eqref{eq:unitarity3}.

\subsection{$N$-body phase space}

In this section, we introduce a general parameterization for the $N$-body phase space integral using a set of spherical variables. 
Following~\cite{Larkoski:2020thc}, the phase space manifold of $N$ massless particles can be decomposed into the product of a $(N-1)$-simplex and a $(2N-3)$-sphere, i.e. 
\begin{equation}
d\Pi_N(2\pi)^4 \ \delta^4\left(P-\sum_{i=1}^Np_i\right)= (2\pi)^{4-3N}s^{N-2} d\Delta_{N-1} \times dS^{2N-3}\ ,
\end{equation}
where $s$ is the center-of-mass energy squared of all the $N$ massless particles. 
In appendix~\ref{sec:appB}, we provide a lightning review on the derivation of the integration measures $d\Delta_{N-1}$ and $dS^{2N-3}$, based on a parametrization using the spinor-helicity variables of the $N$ massless particles involved in the phase space. In the following, we only highlight the main results.

Based on the spinor-helicity formalism, the integration variables can be chosen to be two complex $N$-component vectors, $\vec{u}=(u_1,u_2,\cdots,u_N)^T$ and $\vec{v}=(v_1,v_2,\cdots,v_N)^T$. Intuitively, each pair of complex numbers $(u_i,v_i)$ are the coefficients expanding the spinor $|p_i\rangle$ for each particle on the two basis spinors $|k_1\rangle$ and $|k_2\rangle$ of momenta of incoming particles, i.e. $|p_i\rangle = u_i |k_1\rangle + v_i|k_2\rangle$.

Parameterizing $\vec{u}=(r_1 e^{-i\phi_1},r_2 e^{-i\phi_2},\cdots,r_N e^{-i\phi_N})^T$, and gauge fixing the $\phi_i=0$ with little group redundancy, the integration over $\vec{u}$ can be shown to be equivalent to the integration on a $(N-1)$-simplex, of which the integration measure is given by:
\begin{eqnarray}
d\Delta_{N-1}= \prod_{i=1}^N r_idr_i \ \delta\left(1-\sum^N_{i=1}r_i^2\right)\;.
\label{eq:veu}
\end{eqnarray}
In the spherical coordinates we can further  parameterize $r_i$ as:
\begin{eqnarray}
    r_{N}&=&\cos\theta_{N-1}\ ,\nonumber\\
    r_{N-1}&=&\sin\theta_{N-1}\cos\theta_{N-2}\ ,\nonumber\\
    r_{N-2}&=&\sin\theta_{N-1}\sin\theta_{N-2}\cos\theta_{N-3}\ ,\nonumber\\
    \vdots\nonumber\\
    r_{2} &=& \sin\theta_{N-1}\dots\sin\theta_{2}\cos\theta_{1}\ ,\nonumber\\
    r_1&=&\sin\theta_{N-1}\dots\sin\theta_{2}\sin\theta_{1}\ . \label{eq:prri}
\end{eqnarray}
Since $r_i>0$, all $\theta_i$ are within $[0,\pi/2]$. It follows that
\begin{eqnarray}
   d\Delta_{N-1}  = \frac{1}{2} \left(\prod_{i=1}^{N-1}\sin^{2i-1}\theta_i \cos\theta_i \right) d\theta_1\dots d\theta_{N-1}.
\end{eqnarray}

Likewise, as detailed in the appendix~\ref{sec:appB}, the integral manifold over $\vec{v}$ is equivalent to $S^{2N-3}$ with the constraint
\begin{eqnarray}
    v_{N} = -\frac{\sum_{i=1}^{N-1} r_i v_i}{r_N}.
    \label{eqn:eq:prvii}
\end{eqnarray}
With a change of integration variables into $v'$ such that $(v_1,v_2,\dots,v_{N-1})^T=O (v'_1,v'_2,\dots,v'_{N-1})^T$, the remaining integration measure related to $\vec{v'}$ is given by 
\begin{eqnarray}
   d^{N-1}v'\ \delta\left(1-\sum_{i=1}^{N-1}|v_i'|^2\right)=\left(\prod_{k=1}^{N-2}\cos\eta_k\sin^{2(N-2-k)+1}\eta_k\right)d\xi_i\dots d\xi_{N-1}d\eta_i\dots d\eta_{N-2}, \nonumber \\ 
\end{eqnarray}
with a complex spherical parameterization of $v'$ in the following form:
\begin{eqnarray}
    v_1'&=&e^{-i\xi_1}\cos\eta_1\ ,\nonumber\\
    v_2'&=&e^{-i\xi_2}\sin\eta_1\cos\eta_2\ ,\nonumber\\
    \vdots\nonumber\\
    v_{N-2}'&=&e^{-i\xi_{N-2}}\sin\eta_1\dots\sin\eta_{N-3}\cos\eta_{N-2}\ ,\nonumber\\
    v_{N-1}'&=&e^{-i\xi_{N-1}}\sin\eta_1\dots\sin\eta_{N-3}\sin\eta_{N-2}\ .
    \label{eq:prvi}
\end{eqnarray} 
Here $\xi_i\in [0,2\pi]$, $\eta_i\in [0,\pi/2]$.
Most importantly, we derive an analytical formula for the real and symmetric transformation matrix $O$ in terms of $r_i$:
\begin{eqnarray}
    O=\mathbbm{1}_{(N-1)\times (N-1)}-\left(\frac{\sqrt{1+\beta}-1}{\beta\sqrt{1+\beta}}\right)\frac{\mathbf{r}^T\mathbf{r}}{r_N^2},\ \ \beta=\frac{|\mathbf{r}|^2}{r_N^2},\label{eq:prmatO}
\end{eqnarray}
with $\mathbf{r}=(r_1,r_2,\dots,r_{N-1})$.

With eq.~\eqref{eq:prri}, \eqref{eqn:eq:prvii}, \eqref{eq:prvi} and \eqref{eq:prmatO}, we can construct the explicit parameterization of $\vec{u}$ and $\vec{v}$ in terms of $\eta_i$, $\xi_i$ and $\theta_i$, and the $\delta$-function imposed by momentum conservation is fully resolved. 
To illustrate the power of the general formalism outlined above, we derive explicit expressions for the 3-body and 4-body phase spaces, and the generalization to the $N$-body case follows straightforwardly. 
\begin{itemize}
\item For instance, we find the integration measure for the 3-body phase space  
\begin{eqnarray}
    &&d\Pi_3(2\pi)^4 \delta^4\left(P-\sum_{i=1}^3p_i\right) \nonumber\\
    =&&\frac{s}{2(2\pi)^5 l!}\sin\theta_1\cos\theta_1\sin^3\theta_2\cos\theta_2\sin\eta_1\cos\eta_1 d\theta_1d\theta_2d\eta_1d\xi_1d\xi_2
\end{eqnarray}
with the parametrization of $\vec{u}$ and $\vec{v}$ given by
\begin{eqnarray}
    && u_1=\sin\theta_2\sin\theta_1,\nonumber\\ 
    && u_2=\cos\theta_1\sin\theta_2,\nonumber\\
    && u_3=\cos\theta_2.
\end{eqnarray}
and
\begin{eqnarray}
    &&v_1=e^{-i\xi_1}\cos\eta_1\left(\cos^2\theta_1+\cos\theta_2\sin^2\theta_1\right)+e^{-i\xi_2}\sin\eta_1(\cos\theta_2-1)\cos\theta_1\sin\theta_1\ ,\nonumber \\
    &&v_2=e^{-i\xi_1}\cos\eta_1(\cos\theta_2-1)\cos\theta_1\sin\theta_1+e^{-i\xi_2}\sin\eta_1\left(\cos\theta_2\cos^2\theta_1+\sin^2\theta_1\right)\ ,\nonumber \\
    &&v_3=-\sin\theta_2(e^{-i\xi_1}\cos\eta_1\sin\theta_1+e^{-i\xi_2}\cos\theta_1\sin\eta_1).
\end{eqnarray}

Notice that we have included the symmetry factor $l!$ for identical particles in $d\Pi_3$.
Another parameterization of the 3-body phase space can be found in e.g.~\cite{EliasMiro:2020tdv}. As a consistency check, we verified that different parameterizations yield the same result. 
\item For the 4-body phase space, we find the integration measure
\begin{eqnarray}
    &&d\Pi_4 (2\pi)^4 \delta^4\left(P-\sum_{i=1}^4p_i\right) \nonumber \\
    = && \frac{s^2}{2(2\pi)^{8}l!}\sin\theta_1\cos\theta_1\sin^3\theta_2\cos\theta_2\sin^5\theta_3\cos\theta_3\cos\eta_1\sin^3\eta_1\cos\eta_2\sin\eta_2 \nonumber \\
    && d\theta_1d\theta_2d\theta_3d\xi_1d\xi_2d\xi_3d\eta_1d\eta_2, 
\end{eqnarray}
with the parametrization of $\vec{u}$ and $\vec{v}$ given by
\begin{eqnarray}
    u_1= && \sin\theta_3\sin\theta_2\sin\theta_1,\nonumber\\
    u_2= && \sin\theta_3\sin\theta_2\cos\theta_1,\nonumber \\
    u_3= &&\sin\theta_3\cos\theta_2,\nonumber\\
    u_4= && \cos\theta_3, 
\end{eqnarray}
and
\begin{eqnarray}
v_1=&&e^{-i \xi_2} \sin \eta_1 \cos \eta_2 (\cos \theta_3 - 1) \sin^2 \theta_2 \sin \theta_1 \cos \theta_1 \nonumber \\
    &&+ e^{-i \xi_3} \sin \eta_1 \sin \eta_2 (\cos \theta_3 - 1) \sin \theta_2 \cos \theta_2 \sin \theta_1\nonumber \\
    &&+ e^{-i \xi_1} \cos \eta_1 \left( \sin^2 \theta_2 \left( \cos \theta_3 \sin^2 \theta_1 + \cos^2 \theta_1 \right) + \cos^2 \theta_2 \right),\nonumber\\
   v_2 =&& e^{-i \xi_2} \sin \eta_1 \cos \eta_2 \left( \sin^2 \theta_2 \left( \cos \theta_3 \cos^2 \theta_1 + \sin^2 \theta_1 \right) + \cos^2 \theta_2 \right) \nonumber \\
   &&+ e^{-i \xi_3} \sin \eta_1 \sin \eta_2 (\cos \theta_3 - 1) \sin \theta_2 \cos \theta_2 \cos \theta_1 \nonumber \\
   &&+ e^{-i \xi_1} \cos \eta_1 (\cos \theta_3 - 1) \sin^2 \theta_2 \sin \theta_1 \cos \theta_1,\nonumber\\
   v_3 =&& e^{-i \xi_2} \sin \eta_1 \cos \eta_2 (\cos \theta_3 - 1) \sin \theta_2 \cos \theta_2 \cos \theta_1\nonumber \\
   &&+ e^{-i \xi_3} \sin \eta_1 \sin \eta_2 \left(\cos \theta_3 \cos^2 \theta_2 + \sin^2 \theta_2 \right)\nonumber \\
   &&+ e^{-i \xi_1} \cos \eta_1 (\cos \theta_3 - 1) \sin \theta_2 \cos \theta_2 \sin \theta_1,\nonumber\\
   v_4 =&&-\sin \theta_3 \left[
\sin \theta_2 \left(
e^{-i \xi_2} \sin \eta_1 \cos \eta_2 \cos \theta_1
+e^{-i \xi_1} \cos \eta_1 \sin \theta_1
\right)\right.\nonumber \\
&&\left. +e^{-i \xi_3} \sin \eta_1 \sin \eta_2 \cos \theta_2
\right].
\end{eqnarray}

\end{itemize}

Given the phase space measure $d\Pi_N$, the interference between two local on-shell partial-wave amplitudes can always be computed analytically, since each term in the amplitude square decomposes into a product of $\cos$ and $\sin$ functions of the integration variables. In the following, we illustrate this in two examples for $d\Pi_3$ and $d\Pi_4$, respectively.

\subsection{Normalization of $2\to 3$ and $2\to 4$ partial-wave amplitudes}

Let us consider an operator of the type $\phi_1\phi_2\phi_3\psi_4\psi^\dagger_5D$, which generates on-shell local amplitudes involving particles of the helicity configuration $(0,0,0,-1/2, 1/2)$.~\footnote{Here the convention is that all the particles are incoming. When converting to a 2-to-3 scattering process with 2 incoming and 3 outgoing particles, the helicity configurations can be $(0,0; 0,-1/2,1/2),\ (0,1/2; 0,0,1/2),\ (0,-1/2; 0,0,-1/2),\ (-1/2,1/2; 0,0,0)$, with the semicolon separating incoming and outgoing particles.}
Using the package \texttt{ABC4EFT}~\cite{Li:2022tec}, one can find two independent J-basis amplitudes with the total angular momentum $J=1$ or $0$ for the scattering process 
\begin{equation}
\phi_1(p_1)\phi_2(p_2)\to \phi_3(-p_3)\psi^\dagger_4(-p_4)\psi_5(-p_5)\ .
\end{equation}
The corresponding J-basis amplitudes are the same by applying a proper cyclic relabeling of the external particles
(4321) to the amplitudes given in eq.~\eqref{eq:ampJ1},
where the momentum conservation $\sum_{i=1}^5p_i=0$ and $p^0_{i=3,4,5}<0$ are considered. 
Analytic continuation to the region with $p_i^0>0$ amounts to flipping the sign of the angle spinors $\lambda^\alpha_{3,4,5}$ while leaving the remaining square spinors unchanged, as detailed in appendix~\ref {sec:appA}.
The normalization of these two amplitudes can be calculated accordingly as  
\begin{eqnarray}
    g^{J=0}_{0,0;0,-1/2,1/2}=\Big\langle B^{J=0}_{0,0;0,-1/2,1/2}, B^{J=0}_{0,0;0,-1/2,1/2}\Big\rangle=\frac{s^3}{3072\pi^3},\nonumber \\
    g^{J=1}_{0,0;0,-1/2,1/2}=\Big\langle B^{J=1}_{0,0;0,-1/2,1/2}, B^{J=1}_{0,0;0,-1/2,1/2}\Big\rangle=\frac{s^3}{1024\pi^3}.\label{eq:ampJ2}
\end{eqnarray}

In cases where there is degeneracy for a fixed $J$, the corresponding basis can be orthogonalized within that subspace at a fixed amplitude dimension. This simplifies the computation of the partial-wave unitarity bounds for a given operator.
Again, we consider the operator of the type $\phi^5D^2$ as an example. The J-basis amplitudes are derived in eq.~\eqref{eq:jbphi51} and ~\eqref{eq:jbphi52}, where we find two amplitudes with $J=1$ and three with $J=0$. Using our phase space parameterization of $d\Pi_3$, the interference metric tensor can be obtained analytically as follows:
\begin{eqnarray}
    \Big\langle B^{J=0, a}_{0,0;0,0,0}, B^{J=0, b}_{0,0;0,0,0}\Big\rangle
    &&=\frac{s^3}{256\pi^3l!}\begin{pmatrix}
        \frac{1}{6}& -\frac{1}{12}& \frac{1}{12}\\
        -\frac{1}{12}& \frac{1}{6}& -\frac{1}{12}\\
        \frac{1}{12}& -\frac{1}{12}& \frac{1}{6}
    \end{pmatrix},\\
    \Big\langle B^{J=1, a}_{0,0;0,0,0}, B^{J=1, b}_{0,0;0,0,0}\Big\rangle
    &&=\frac{s^3}{256\pi^3l!}\begin{pmatrix}
        \frac{1}{6}& \frac{1}{12}\\
        \frac{1}{12}& \frac{1}{6}
    \end{pmatrix},
\end{eqnarray}
here we still leave the general symmetric factor $l!$ in the denominator, though two particles can be differed by gauge quantum numbers. 
From this result, one can obtain the orthogonal basis and determine the corresponding normalization factors: 
\begin{eqnarray}
   & B^{'J=0, a=1,2,3}_{0,0;0,0,0}: 
    \begin{cases}
        \frac{s_{34}+s_{35}+s_{45}}{\sqrt{3}} &g=\frac{s^3}{768 \pi^3l!},\\
        \frac{-s_{34}+s_{45}}{\sqrt{2}} &g=\frac{s^3}{3072 \pi^3l!},\\
        \frac{s_{34}-2s_{35}+s_{45}}{\sqrt{6}} &g=\frac{s^3}{3072 \pi^3l!},
    \end{cases}\\
   & B^{'J=1, a=1,2}_{0,0;0,0,0}:\begin{cases}
        \frac{2s_{24}-2s_{25}+s_{34}-s_{35}}{\sqrt{2}} & g=\frac{s^3}{1024\pi^3l!},\\
        \frac{2s_{24}+2s_{25}+s_{34}+s_{35}+2s_{45}}{\sqrt{2}} & g=\frac{s^3}{3072\pi^3l!}.
    \end{cases}
\end{eqnarray}

Finally, let us consider a 2-to-4 scattering process 
\begin{equation}
\phi_1(p_1)\phi_2(p_2)\to \phi_3(-p_3)\phi_4(-p_4)\psi^\dagger(-p_5)\psi(-p_6) .
\end{equation}
Again, we first obtain the corresponding J-basis amplitudes 
\begin{eqnarray}
    &&B^{J=1} = 2\ab{25}[26]+\ab{35}[36]+\ab{45}[46],\label{eq:jbasis1}\\
    &&B^{J=0}_1 = \ab{35}[36],\quad 
    B^{J=0}_2 = \ab{45}[46],
\end{eqnarray}
where we find two amplitudes in the $J=0$ subspace. Following the same procedure of integrating over $d\Pi_4$, we have the normalization factors and the corresponding orthogonal J-basis amplitudes. For $J=0$, we have
\begin{eqnarray}
   B'^{J=0}_{1} = \frac{\ab{35}[36]+\ab{45}[46]}{\sqrt{2}}, \quad g^{J=0}_1 =\frac{s^4}{737280\pi^5}, \label{eq:jbasis2}\\
   B'^{J=0}_{2} = \frac{-\ab{35}[36]+\ab{45}[46]}{\sqrt{2}}, \quad g^{J=0}_2=\frac{s^4}{1474560\pi^5},\label{eq:jbasis3}
\end{eqnarray}
while for the $J=1$ case we have
\begin{eqnarray}
    g^{J=1} = \frac{s^4}{184320\pi^5}.
\end{eqnarray}

\subsection{A Mathematica code implementing phase space integration}

To assist the reader in analytically computing the phase space integrals for the interference between partial-wave amplitudes, we provide the Mathematica code, which can be downloaded in from the GitHub link\footnote{\url{https://github.com/haolinli1991/CalcJMetric}}. 

The key function is \mmaInlineCell[pattern={amp1,amp2,n,in},defined=PSIntAMPUser]{Code}{PSIntAMPUser[amp1,amp2,n,in]}.
It takes four inputs: the first two are the amplitudes whose interference is being calculated; the third is the total number of final-state particles $n$; the fourth is a list of two indices indicating which two particles are treated as the initial-state ones. 
Overall, the function evaluates the following integral:
\begin{eqnarray}
   {\rm PSIntAMPUser[M_1, M_2,n,\{i,j\}]} =\int d\Pi_{k\notin \{i,j\}}(2\pi)^4\delta^4(p_i+p_j-\sum_{k\notin \{i,j\}}p_k){\rm M^*_1}{\rm M_2}.\nonumber \\
\end{eqnarray}
As an example, the normalization factor $g^{J=0}_1$ in the eq.~\eqref{eq:jbasis2} can be computed easily as follows,
\begin{mmaCell}[moredefined={PSIntAMPUser,ab,sb}]{Input}
  PSIntAMPUser[\mmaFrac{ab[3,5]sb[3,6]+ab[4,5]sb[4,6]}{\mmaSqrt{2}}, \\ \mmaFrac{ab[3,5]sb[3,6]+ab[4,5]sb[4,6]}{\mmaSqrt{2}},4,\{1,2\}]
\end{mmaCell}
\begin{mmaCell}{Output}
  \mmaFrac{\mmaSup{s}{4}}{737280\mmaSup{\(\pi\)}{5}}
\end{mmaCell}
where \mmaInlineCell[moredefined=ab]{Code}{ab} and \mmaInlineCell[moredefined=sb]{Code}{sb} denote the angle and square brackets, and all momenta are treated as incoming. The variable \texttt{s} in the output represents the center-of-mass energy squared of the two initial-state particles. We validate our calculation against the known total volume of the three- and four-body phase space and verify across multiple examples that partial-wave amplitudes with different total angular momentum are indeed orthogonal. 

\begin{figure}[h]
    \centering
    \includegraphics[width=0.75\textwidth]{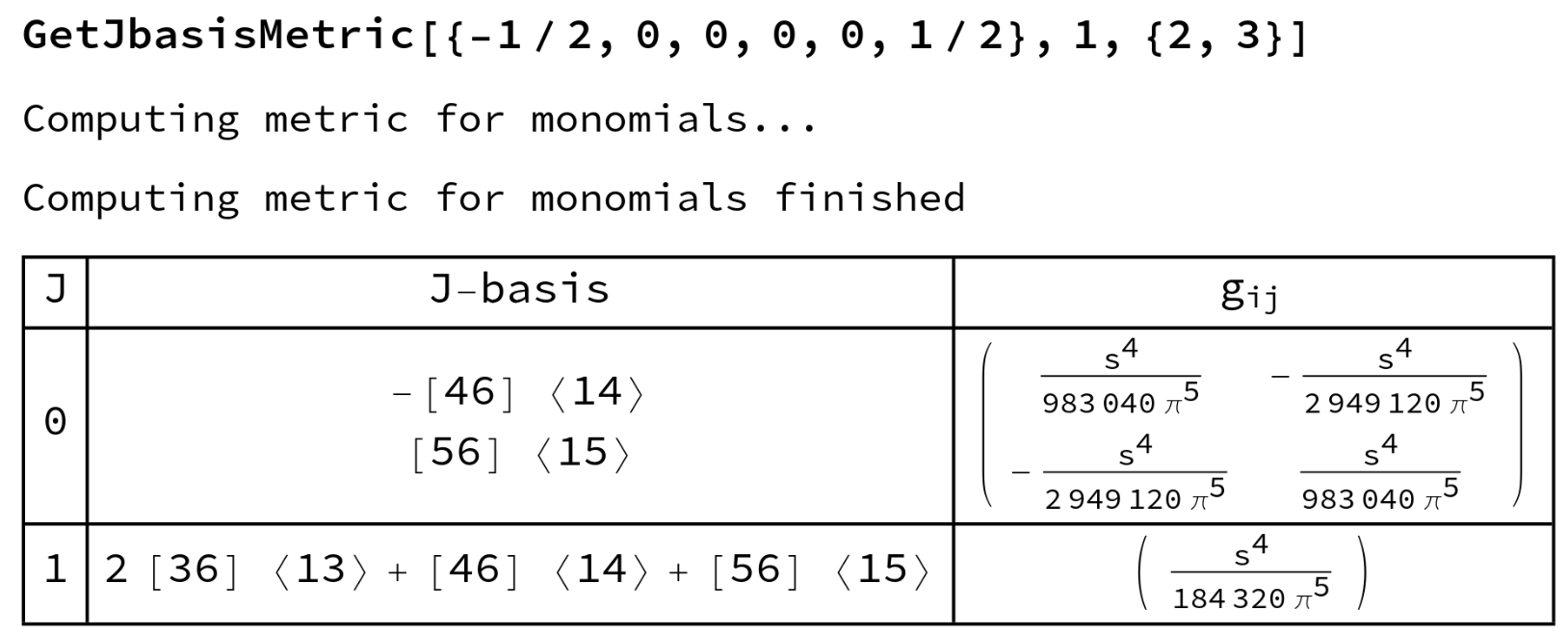}
    \caption{An example output of the function   \texttt{GetJbasisMetric}.}
    \label{fig:GJM}
\end{figure}

Furthermore, we also provide the function to compute the metric of the J-basis amplitudes, given the helicities of the external particles and the number of derivatives in the corresponding operators. 
The function \mmaInlineCell[pattern={hellist,nD,in},defined=GetJbasisMetric]{Code}{GetJbasisMetric[hellist,nD,in]} takes three inputs: the first is the list of helicities of external particles $\{h_i\}$ ordered from the most negative to the most positive; the second parameter represents the number of derivatives $\# D$ in the corresponding operator, which fixes the mass dimension of the amplitude to be $\sum_{i=1}^n |h_i|+\# D$, where $n$ is the total number of external particles; the last input is the label of particles that are considered to be the incoming particles. 
The output of this function is a table as in figure.~\ref{fig:GJM}, where the first column includes the possible values of total angular momentum for the scattering channel; the second column includes the J-basis amplitudes obtained from the \mmaInlineCell[pattern={k,ch,state},defined=W2Diagonalize]{Code}{W2Diagonalize};
the last column includes the metrics for the inner product of the amplitudes in the same $J$ subspace, and the labeling of the indices is in order with the ones in the second column.

\section{Partial-wave unitarity bounds for dimension-7 and dimension-8 operators in SMEFT}\label{sec:smeft_example}

In this section, we use the orthogonal partial-wave amplitude basis, obtained in the previous section, to derive the perturbative unitarity bounds for two SMEFT operators, i.e. $\overline{e_R} \overline{l_L}H^{\dagger3}D$ and $H^2H^{\dagger2}\overline{e_R}e_R D$. In particular, we will illustrate how to treat the gauge and flavor quantum numbers when deriving the bounds.

\subsection{A dimension-7 example: $\overline{e_R} \overline{l_L}H^{\dagger3}D$}

We first consider an operator that violates the lepton number and can contribute to the process of neutrinoless double beta decay~\cite{Liao:2019tep}. At the dimension-7 level, there is only one independent operator of this type.
In four-component fermion notation, the operator is given by
\begin{eqnarray}
    {\cal O}_1=iC_{f_1f_5}\epsilon_{i_2i_4}\epsilon_{i_3i_5}H^{\dagger i_2}H^{\dagger i_3}\left( D_\mu H^{\dagger i_4}\right)\left(\overline{e_R}\gamma^\mu {\cal C} \overline{l^{i_5}_L}^{T}\right),
\end{eqnarray}
where $i_{2,3,4,5}$ are $SU(2)_L$ gauge indices, $C_{f_1f_5}$ is the Wilson coefficient with $f_{1,5}$ being flavor indices of the right-handed electron $e_R$ and the left-handed lepton doublet $l_L$. 
The charge conjugation operator ${\cal C}$ is defined as ${\cal C}=i\gamma^0\gamma^2$. Summation over the repeated indices is understood. 

The operator contributes to the scattering process 
\begin{equation}
H^{\dagger i_2}(p_2) H^{\dagger i_3}(p_3)\to e^{-}_{f_1}(-p_1) H^{i_4}(-p_4)L^{i_5}_{f_5}(-p_5) \;, 
\end{equation}
with the corresponding amplitude:
\begin{eqnarray}
    M_{i_2i_3i_4i_5}^{f_1f_5}=C_{f_1f_5}\epsilon_{i_2i_4}\epsilon_{i_3i_5}[45]\ab{14}+{\rm sym (234)},\label{eq:ampsym}
\end{eqnarray}
where symmetrization applies to both the momenta and the gauge indices.
The Lorentz J-basis amplitudes are obtained as in eq.~\eqref{eq:ampJ1}
\begin{eqnarray}
    B^{J=0} = \ab{14}[45], \quad B^{J=1} = -(2\ab{13}[35]+\ab{14}[45]),
\end{eqnarray}
and their normalization factors can be found in eq.~\eqref{eq:ampJ2}.
One can find the coordinates of the full amplitude $M_{i_2i_3i_4i_5}^{f_1f_5}$ on this J-basis as
\begin{eqnarray}
    &&C^{J=1} = \frac{C_{f_1f_5}}{2}\left(  \epsilon_{i_2i_4} \epsilon_{i_3i_5}+2 \epsilon_{i_2i_3}\epsilon_{i_4i_5} -\epsilon_{i_2i_5} \epsilon_{i_3i_4}\right),\\
    &&C^{J=0} = \frac{3C_{f_1f_5}}{2}\left( \epsilon_{i_2i_5} \epsilon_{i_3i_4}+ \epsilon_{i_2i_4} \epsilon_{i_3i_5}\right).\
\end{eqnarray}
By fixing the $SU(2)$ indices, we obtain the coefficients of partial-wave decomposition shown in table.~\ref{tab:pwd1}. 
\begin{table}[]
    \centering
    \begin{tabular}{|c|c|c|c|c|c|}
    \hline
        $i_2$ & $i_3$ & $i_4$ & $i_5$ & $C^{J=1}$ & $C^{J=0}$ \\
        \hline
         1 & 1 &  2 & 2 & 0 & $3C_{f_1f_2}$\\
         \hline
         \multirow{2}{*}{1} & \multirow{2}{*}{2} &  1 & 2 & $-3C_{f_1f_5}/2$ & $-3C_{f_1f_5}/2$ \\
         \cline{3-6}
         & & 2 & 1 & $3C_{f_1f_5}/2$ & $-3C_{f_1f_5}/2$ \\
         \hline
         \multirow{2}{*}{2} & \multirow{2}{*}{1} &  1 & 2 & $3C_{f_1f_5}/2$ & $-3C_{f_1f_5}/2$ \\
         \cline{3-6}
         & & 2 & 1 & $-3C_{f_1f_5}/2$ & $-3C_{f_1f_5}/2$ \\
         \hline
         2 & 2 &  1 & 1 & 0 & $3C_{f_1f_2}$\\
         \hline
    \end{tabular}
    \caption{Partial-wave decomposition of the amplitude in eq.~\eqref{eq:ampsym} for the scattering channel $H^\dagger H^\dagger \to e^- HL$, where $C^{J=0}$ and $C^{J=1}$ are the coordinates of the amplitude projecting to the j-basis amplitudes $B^{J=0}$ and $B^{J=1}$.}
    \label{tab:pwd1}
\end{table}
The resulting partial-wave unitarity bounds on $C_{f_1f_5}$ for different choices of $i_2$ and $i_3$ (while $i_{4,5}$ have been summed over) are
\begin{eqnarray}
    &i_2=i_3=1 \text{ or } i_2=i_3=2,& \quad \frac{3 s^3\sum_{f_1f_5}|C_{f_1f_5}|^2}{8192\pi^4} \leq 1 \text{ for } J=0 ,\label{eq:bound1}\\
    \nonumber \\
    &i_2=1,\ i_3=2 \text{ or } i_2=2,\ i_3=1,\quad  &\begin{cases}
        \dfrac{ s^3\sum_{f_1f_5}|C_{f_1f_5}|^2}{8192\pi^4} \leq 1 , & J=1\\
        
         \dfrac{3 s^3\sum_{f_1f_5}|C_{f_1f_5}|^2}{16384\pi^4} \leq 1 , & J=0
    \end{cases} .
\end{eqnarray}
Hence, we find that eq.~\eqref{eq:bound1} gives the strongest bound.

The same operator also contributes to the scattering channel 
\begin{equation}
e^{+}_{f_1}(p_1)  H^{\dagger i_2}(p_2)\to  H^{i_3}(-p_3)H^{i_4}(-p_4)L^{i_5}_{f_5}(-p_5) \;.
\end{equation}
For this channel, there are two degenerate J-basis amplitudes with $J=1/2$. The orthogonal J-basis amplitudes and their normalization factors are 
\begin{eqnarray}
    &B^{J=1/2}_1 = \frac{\ab{13}[35]+\ab{14}[45]}{\sqrt{2}}, &g_1^{J=1/2} =  \frac{s^3}{1536\pi^3l!},\label{eq:pwb1}\\
    &B^{J=1/2}_2 = \frac{-\ab{13}[35]+\ab{14}[45]}{\sqrt{2}}, &g_2^{J=1/2} =  \frac{s^3}{3072\pi^3l!}.\label{eq:pwb2}
\end{eqnarray}
The coefficients of the full amplitude projected on the J-basis amplitudes are
\begin{eqnarray}
    &&C^{J=1/2}_1= \frac{3C_{f_1f_5}}{\sqrt{2}}\left( \epsilon_{i_2i_4}\epsilon_{i_3i_5}+\epsilon_{i_2i_3}\epsilon_{i_4i_5}\right),\\
    &&C^{J=1/2}_2 = \frac{3 C_{f_1f_5}}{\sqrt{2}}\left(\epsilon_{i_2i_5}\epsilon_{i_3i_4}\right)\;.
\end{eqnarray}
In table.~\ref{tab:pwd2}, we summarize the values of $C^{J=1/2}_{1,2}$ for different choices of the $SU(2)_L$ gauge indices.
\begin{table}[]
    \centering
    \begin{tabular}{|c|c|c|c|c|c|}
    \hline
        $i_2$ & $i_3$ & $i_4$ & $i_5$ & $C^{J=1/2}_1$ & $C^{J=1/2}_2$ \\
        \hline
         \multirow{3}{*}{1} & 1 &  2 & 2 & $3C_{f_1f_5}/\sqrt{2}$ & $3C_{f_1f_5}/\sqrt{2}$\\
         \cline{2-6}
          & 2 &  1 & 2 & $3C_{f_1f_5}/\sqrt{2}$ & $-3C_{f_1f_5}/\sqrt{2}$ \\
         \cline{2-6}
         & 2& 2 & 1 & $-3\sqrt{2}C_{f_1f_5} $ & $0$ \\
         \hline
         \multirow{3}{*}{2} & 1 &  1 & 2 & $-3\sqrt{2}C_{f_1f_5} $ & $0$ \\
         \cline{2-6}
         & 1& 2 & 1 & $3C_{f_1f_5}/\sqrt{2}$ & $-3C_{f_1f_5}/\sqrt{2}$ \\
         \cline{2-6}
          & 2 &  1 & 1 & $3C_{f_1f_5}/\sqrt{2}$ & $3C_{f_1f_2}/\sqrt{2}$\\
         \hline
    \end{tabular}
    \caption{Partial-wave decomposition of the amplitude in eq.~\eqref{eq:ampsym} for the scattering channel $e^+ H^\dagger \to HHL$, where $C^{J=1/2}_1$ and $C^{J=1/2}_2$ are the coordinates of the amplitude projecting to the J-basis amplitudes $B^{J=1/2}_1$ and $B^{J=1/2}_2$.}
    \label{tab:pwd2}
\end{table}
Considering the initial state where three flavors of $e^{+}$ mix evenly, we obtain the most stringent bounds on the Wilson coefficients as follows: 
\begin{eqnarray}
   i_2=1\ {\rm or}\ 2,\quad \dfrac{3 s^3\sum_{f_1f_5}|C_{f_1f_5}|^2}{8192} \leq 1,\label{eq:bound3}
\end{eqnarray}
where we consider the symmetry factor in the phase-space integral for the two identical scalars in the final states when $i_3=i_4$.
For the scattering channel 
\begin{equation}
L^{\dagger i_5}_{f_5}(p_5)   H^{\dagger i_2}(p_2)\to e^{-}_{f_1}(-p_1)  H^{ i_3}(-p_3)H^{i_4}(-p_4)\;, 
\end{equation}
the partial-wave amplitude and the decomposition are the same as the scattering channel with $e^+ H^\dagger$ as the initial state. Therefore, it leads to the same bound in eq.~\eqref{eq:bound3}.

Finally, we investigate the scattering channel 
\begin{equation}
e^{-}_{f_1}(p_1) L^{\dagger i_5}_{f_5}(p_5)   \to  H^{\dagger i_2}(-p_2) H^{ i_3}(-p_3)H^{i_4}(-p_4)\;. 
\end{equation}
We find that the partial-wave amplitudes have the same form as in eq.~\eqref{eq:pwb1} and eq.~\eqref{eq:pwb2}, even though they correspond to total angular momentum $J=1$ instead of $J=1/2$. The normalization factors change correspondingly, since they have different helicity configurations in the final state, 
\begin{eqnarray}
    &B^{J=1}_1 = \frac{\ab{13}[35]+\ab{14}[45]}{\sqrt{2}}, &g_1^{J=1} =  \frac{s^3}{6144\pi^3l!},\\
    &B^{J=1}_2 = \frac{-\ab{13}[35]+\ab{14}[45]}{\sqrt{2}}, &g_2^{J=1} =  \frac{s^3}{2048\pi^3l!}.
\end{eqnarray}
Since the form of the partial wave amplitudes is the same as in the previous scattering channel, the decomposition coefficients are also the same as in table.~\ref{tab:pwd2}, from which one can obtain the bounds on each flavor component of the Wilson coefficients:
\begin{eqnarray}
   \frac{5|C_{f_1f_5}|^2s^3}{32768 \pi^4}  \leq 1.
\end{eqnarray}
Comparing bounds derived from different channels, we can find that the strongest bound is given by the $J=0$ partial wave in the scattering $H^\dagger H^\dagger \to e^- H L$ as in eq.~\eqref{eq:bound1} and $J=1/2$ partial wave in the scattering $e^+ H^\dagger \to e^- HH L$.

\subsection{A dimension-8 example: $H^2H^{\dagger2}\overline{e_R}e_R D$}

At last, we consider a dimension-8 operator
\begin{eqnarray}
    {\cal O}_2 = |H|^2H^\dagger \overleftrightarrow{D}_\mu H(\overline{e_R}_{f_6}\gamma^\mu e_{Rf_1}),
\end{eqnarray}
where $e_R$ is the right-handed electron.
The contact on-shell amplitude induced by this operator is given by
\begin{eqnarray}
    M_{i_2i_3i_4i_5}^{f_1f_6}=&&C_{f_1f_6}\Big\lbrace \left(\delta_{i_4}^{i_2}\delta_{i_5}^{i_3}\ab{15}[56]+{\rm sym}(45)\right)+{\rm sym}(23) \nonumber \\
    -&& \left(\delta_{i_4}^{i_2}\delta_{i_5}^{i_3}\ab{13}[36]+{\rm sym}(23)\right)+{\rm sym}(45)\Big\rbrace,\label{eq:M2}
\end{eqnarray}
where $p_1$ to $p_6$ in order represent the momenta for lepton, two Higgs, two anti-Higgs, and one anti-lepton respectively in the all incoming convention, $i_{2,3}$ and $i_{4,5}$ represent the $SU(2)_L$ indices for Higgs and anti-Higgs respectively, $f_1$ and $f_6$ are the flavor for lepton and anti-lepton.
The J-basis for the scattering channel $H_{i_2}(p_2)H_{i_3}(p_3)\to e^+(-p_1)e^-(-p_6)H^{\dagger i_4}(-p_4)H^{\dagger i_5}(-p_5)$ give:
\begin{eqnarray}
    B^{J=1}&&=2\ab{13}[36]+\ab{14}[46]+\ab{15}[56],\quad g^{J=1}=\frac{s^4}{184320\pi^5}\label{eq:ampj1}\\
    B^{J=0}_1&& = \frac{\ab{14}[46]+\ab{15}[56]}{\sqrt{2}},\quad g^{J=1}_1=\frac{s^4}{737280\pi^5},\\
    B^{J=0}_2&& = \frac{-\ab{14}[46]+\ab{15}[56]}{\sqrt{2}},\quad g^{J=1}_2=\frac{s^4}{1474560\pi^5},\label{eq:ampj3}
\end{eqnarray}
as one expected, the orthogonal basis is equivalent to the ones in eq.~\eqref{eq:jbasis1} to eq.~\eqref{eq:jbasis3} by cyclic permutation among (12345). The full amplitude eq.~\eqref{eq:M2} can be expanded by the J-basis, and we find that only the coefficient of $B^{J=0}_1$ is non-zero:
\begin{eqnarray}
    C^{J=0}_1 = 2\sqrt{2}(\delta_{i_4}^{i_2}\delta_{i_5}^{i_3}+\delta_{i_4}^{i_3}\delta_{i_5}^{i_2})C_{f_1f_6}.
\end{eqnarray}
The tightest bound for this channel is given by setting $i_2=1$ and $i_3=2$ or vice versa as $i_2=i_3$ indicates $i_4=i_5$ in the final state, introducing a factor of $2!$ in the denominator of $g^{J=0}_1$ due to the identical particle in the final state:
\begin{eqnarray}
    \frac{\sum_{f_1f_5}|C_{f_1f_6}|^2 s^4}{737280\pi^6}\leq 1.
\end{eqnarray}

Similarly, we obtain the optimal bounds for the rest of the possible scattering channels, and they are
\begin{eqnarray}
     &&\frac{\sum_{f_1f_6}|C_{f_1f_6}|^2 s^4}{737280\pi^6}\leq 1, {\rm for }\ H_{i_2}(p_2)H^{\dagger i_4}(p_4),\ H^{\dagger i_4}(p_4)H^{\dagger i_5}(p_5),\\
     &&\frac{\sum_{f_6}|C_{f_1f_6}|^2 s^4}{737280\pi^6}\leq 1, {\rm for }\ H_{i_2}(p_2)e^-(p_1),\ H^{\dagger i_4}(p_4)e^-(p_1),\\
     &&\frac{\sum_{f_1}|C_{f_1f_6}|^2 s^4}{737280\pi^6}\leq 1, {\rm for }\ H_{i_2}(p_2)e^+(p_6),\ H^{\dagger i_4}(p_4)e^+(p_6).
\end{eqnarray}
If we assume a $U(3)_{e_R}$ flavor symmetry, then only the diagonal Wilson coefficients are non-zero and are equal to the same value, then all the different scattering channels provide the same bound on the Wilson coefficient.

To summarize, to obtain the optimal partial-wave unitarity bound for a given SMEFT operator containing more than four fields, one proceeds with the following steps:
\begin{enumerate}
    \item Construct the amplitude basis for the type of operator being considered.
    \item Determine the partial-wave (i.e. J-basis) amplitudes using the Casimir operator method.
    \item Identify all possible 2-to-$N$ scattering channels relevant to the operator.
    \item For each scattering channel, decompose the amplitude onto the J-basis, and derive the strongest unitarity bound among different partial waves and gauge components.
    \item Iterate over all the scattering channels and select the optimal bound as the final result.
\end{enumerate}
Note that in the examples discussed above, we have, for simplicity, assumed that only a single SMEFT operator contributes in the analysis. However, in a complete analysis, multiple operators can contribute to the same scattering channel. We leave such a comprehensive analysis to future work. 

\section{Conclusion}
\label{sec:conc}

In this paper, we study the generalized partial-wave unitarity bounds derived from 2-to-$N$ scattering amplitudes and show that they impose theoretical consistency constraints on the Wilson coefficients of higher-dimensional operators in effective field theories involving more than four fields.

We outline a general procedure --- based on the Casimir operator method --- for systematically constructing the local on-shell partial-wave amplitude basis at a given mass dimension.
This method uses the Young tensor method to generate an amplitude basis, from which the representation matrix of the $W^2$ Casimir operator of the Poincar\'e group is constructed. The partial-wave amplitude basis is then obtained by diagonalizing this matrix.
We further highlight that, in general 2-to-$N$ scattering processes, partial-wave amplitudes for a fixed total angular momentum and a given mass dimension can be degenerate. Such a degeneracy can be partially resolved by considering mutually commuting angular momentum operators associated with subsets of the final-state particles.

To determine the unitarity bound for each partial wave, it is first necessary to compute the normalization factor. 
This requires computing the $N$-body phase space integrals between pairs of amplitudes in the partial-wave basis.
To enable analytic evaluation of the massless $N$-body phase space integrals, we map the integral onto a compact manifold, given by the product of a $(N-1)$-simplex and a $(2N-3)$-sphere.  
We show that the coordinate transformation between the four-momenta of final-state particles and the spherical coordinates on this compact manifold can be solved analytically by leveraging the spinor-helicity representation of massless momenta.
As examples, we provide concrete parameterizations of the spinor-helicity variables for the 3- and 4-body final states using the aforementioned spherical coordinates, along with the corresponding integration measures.
Additionally, we provide Mathematica code that automates the analytical computation of the resulting phase space integrals. 
Finally, with two examples, we illustrate how to derive the partial-wave unitarity bounds on the Wilson coefficients of SMEFT operators containing more than four fields.

We believe that these theoretical bounds will serve as valuable inputs when applying effective field theory to interpret results from precision measurements across a wide range of current and future experimental programs. 

\begin{acknowledgments}
H.-L.L. Thanks Gauthier Durieux, Ming-Lei Xiao, Jiang-Hao Yu and Yu-Hui Zheng's helpful discussion.
The work of H.-L.L. si supported by the start-up funding of Sun Yat-Sen University under grant number 74130-12255013 and by the National Science Foundation of China under Grants No.1250050417.
The work of L.X.X. is partially supported by "Exotic High Energy Phenomenology" (X-HEP)~\footnote{Funded by the European Union. Views and opinions expressed are however those of the author(s) only and do not necessarily reflect those of the European Union or the ERC Executive Agency (ERCEA). Neither the European Union nor the granting authority can be held responsible for them.}, a project funded by the European Union - Grant Agreement n.101039756. L.X.X. is also supported by the Munich Institute for Astro-, Particle and BioPhysics (MIAPbP) which is funded by the Deutsche Forschungsgemeinschaft (DFG, German Research Foundation) under Germany's Excellence Strategy – EXC-2094 – 390783311.
\end{acknowledgments}

\appendix
\section{Conventions for spinor-helicity variables}
\label{sec:appA}

We adopt the following parameterization for the massless spinor-helicity variables associated with positive-energy momenta $E_i>0$, where $i$ labels the $i$-th external particle:
\begin{eqnarray}
\langle i|^\alpha\equiv \lambda^\alpha_i=\sqrt{2E_i}\begin{pmatrix}
\cos\theta_i/2\\
e^{-i\phi_i}\sin\theta_i/2
\end{pmatrix},\quad
[\tilde{i}|_{\dot\alpha}\equiv \tilde{\lambda}_{i\dot\alpha}=\sqrt{2E_i}\begin{pmatrix}
-e^{i\phi_i}\sin\theta_i/2\\
\cos\theta_i/2
\end{pmatrix}.\label{eq:defspinor}
\end{eqnarray}
The spinor indices can be raised and lowered using the totally anti-symmetric $\epsilon$ tensors defined by $\epsilon^{12}=\epsilon_{21}=\epsilon^{\dot{1}\dot{2}}=\epsilon_{\dot{2}\dot{1}}=1$, so that 
\begin{eqnarray}
|i\rangle_{\alpha}\equiv\lambda_{i\alpha}=\epsilon_{\alpha\beta}\lambda^{\beta}_i,\quad 
|\tilde{i}]^{\dot\alpha}\equiv\tilde{\lambda}^{\dot\alpha}_i=\epsilon^{\dot\alpha\dot\beta}\tilde{\lambda}_{i\dot\beta}.
\end{eqnarray}
One can verify that this parameterization indeed satisfies the following identities:
\begin{eqnarray}\label{eq:pdef}
    p_{i\alpha\dot\alpha}\equiv p_{i\mu}\sigma^{\mu}_{\alpha\dot\alpha} =\lambda_{i\alpha}\tilde{\lambda}_{i\dot\alpha} ,\quad  p_{i}^{\dot\alpha\alpha}\equiv p_{i\mu}\bar{\sigma}^{\mu\dot\alpha\alpha} =\tilde{\lambda}^{i\dot\alpha}\lambda^{i\alpha}.
\end{eqnarray}
where $p^{\mu}=(E,p_x, p_y,p_z)$, $\sigma^{\mu}=(\mathbbm{1},\boldsymbol{\sigma})$, $\bar{\sigma}^{\mu}=(\mathbbm{1},-\boldsymbol{\sigma})$, with $\boldsymbol{\sigma}$ denoting the Pauli matrices.

A state vector $|\mathbf{p};h,M=0\rangle$ in the massless unitary irreducible representation of the Poincar\'e group is labeled by the three momentum vector $\mathbf{p}$, which corresponds to the eigenvalue of the spatial translation generator. On the other hand, the helicity $h$ labels the irreducible representation of the little group (i.e. the two-dimensional Euclidean group), which underlies the construction of massless Poincar\'e representations. From Wigner's little group classification, a general Lorentz transformation $\Lambda$ acting on the state $|\mathbf{p};h,M=0\rangle$ induces a phase factor that depends on both $\Lambda$ and the momentum $p$ of the state. One can view the spinor variables as a realization of the massless irreducible representation of the Poincar\'e group with $h=\pm 1/2$, such that
\begin{eqnarray}
    &&\Lambda\to D^{-\frac{1}{2}}[\Lambda],\quad |\mathbf{p};h\rangle \to \lambda(p),\\
    &&D^{-\frac{1}{2}}[\Lambda].\lambda(p) = e^{iw(\Lambda,p)/2}\lambda(\Lambda p),
\end{eqnarray}
where $D^{-\frac{1}{2}}[\Lambda]$ denotes the spin-$1/2$ representation of the Lorentz group and $w(\Lambda,p)$ is the little group phase factor.

The parameterization of the spinors in eq.~\eqref{eq:defspinor} is defined for momentum with $p^0>0$. For negative energy momenta (i.e., $p^0<0$), we analytically extend the definition of the spinor variables as follows:
\begin{eqnarray}
    \lambda_\alpha(p)\equiv -\lambda_\alpha(-p),\ \tilde{\lambda}_{\dot\alpha}(p)\equiv \tilde{\lambda}_{\dot\alpha}(-p),\text{ for $p^0<0$},   
\end{eqnarray}
which preserves eq.~\eqref{eq:pdef}. In addition, for real momenta, the spinor satisfy $(\lambda_{\alpha})^*=\pm \tilde{\lambda}_{\dot\alpha}$, where the minus sign applies to momenta with $p^0<0$.

\section{Phase space integral for $N$ massless particles}
\label{sec:appB}

We review the $N$-body phase space of massless particles following~\cite{Larkoski:2020thc}. We leave the more technical discussions in this appendix while highlighting our main results in section~\ref{sec:phase-space}. For a more complete discussion, we refer the readers to~\cite{Larkoski:2020thc}.

The standard $N$-body phase space is given by
\begin{eqnarray}
    d\Pi_N (2\pi)^4 \delta^4\left(P-\sum_{i=1}^Np_i\right)&=& \prod_{i=1}^N\frac{d^3p_i}{(2\pi)^3 2E_i} \ (2\pi)^4 \delta^4\left(P-\sum_{i=1}^Np_i\right)\\
    &=& (2\pi)^{4-3N} \prod_{i=1}^N d^4 p_i \delta(p_i^2)\Theta(E_i) \ \delta^4\left(P-\sum_{i=1}^Np_i\right), 
\end{eqnarray}
where $s=P^2$ is the square of the center-of-mass energy.
When all particles are massless, instead of using the three-momenta $p_i$, it is convenient to use the spinor-helicity variables as the integration variables, where for each four-momentum we replace $p_i$ with $(p_i)_{\alpha\dot\alpha}=\lambda_{i\alpha}\tilde{\lambda}_{i\dot\alpha}$ and $(\tilde{\lambda}_{i\dot\alpha})^*=\lambda_{i\alpha}$. Hence, the integration measure for the on-shell four-momentum $p_i$ becomes
\begin{equation}
d^4 p_i \delta(p_i^2)\Theta(E_i)=\frac{d^2 \lambda_{i\alpha} \ d^2 \tilde{\lambda}_{i\dot\alpha}}{U(1)}\equiv s \frac{d u_i \ d v_i}{U(1)} \ ,
\end{equation}
where the $U(1)$ in the denominator denotes the redundant little group transformation (that should eventually be removed for each $p_i$), $u_i$ and $v_i$ are the two complex numbers parameterizing $p_i$. One can view $u_i$ and $v_i$ as the coefficients when projecting the spinor $|p_i\rangle$ on the two basis spinors $|k_1\rangle$ and $|k_2\rangle$, i.e. $|p_i\rangle = u_i |k_1\rangle + v_i|k_2\rangle$.~\footnote{One convenient choice for the basis spinors is that $|k_1\rangle=(1,0)^T$ and $|k_2\rangle=(0,1)^T$, which can be interpreted as the two spinors of the initial-state particles moving in the forward and backward directions, respectively.} All the $N$ pairs of the two complex numbers $(u_i, v_i)$ form two complex $N$-component vectors $\vec{u}=(u_1,u_2,\cdots,u_N)^T$ and $\vec{v}=(v_1,v_2,\cdots,v_N)^T$, i.e.,
\begin{eqnarray}
    \begin{pmatrix}
        |p_1\rangle &
        |p_2\rangle &
        \cdots &
        |p_N\rangle
    \end{pmatrix}^T=
    \begin{pmatrix}
        u_1 & u_2& \cdots & u_N\\
        v_1 & v_2& \cdots & v_N\\
    \end{pmatrix}^T
    \begin{pmatrix}
        |k_1\rangle \\
        |k_2\rangle
    \end{pmatrix}.
\end{eqnarray}
With the parameterization of $u_i$ and $v_i$, the phase space can be written as
\begin{eqnarray}
    d\Pi_N=(2\pi)^{4-3N}s^{N-2}\frac{d^N u d^N v}{U(1)^N} \delta(1-|\vec{u}|^2)\delta(1-|\vec{v}|^2)\delta^{2}(\vec{u}^\dagger \vec{v}),
\end{eqnarray}
where the superscript $2$ in the last $\delta$-function indicates that it is for a complex variable. More precisely, for $P_{\alpha \dot \alpha}-\sum_{i=1}^N(p_i)_{\alpha \dot \alpha}$ inside the $\delta$ function, $1-|\vec{u}|^2$ matches the $(\alpha,\dot \alpha)=(1,1)$ component, $1-|\vec{v}|^2$ matches the $(\alpha,\dot \alpha)=(2,2)$ component, while $\vec{u}^\dagger \vec{v}$ matches the $(\alpha,\dot \alpha)=(1,2)$ component.

More specifically, removing the redundancy of the little group $U(1)^N$ factor amounts to fixing the phase of one of the complex variables for each pair of $u_i$ and $v_i$. Without loss of generality, we choose to set $u_i=r_i e^{i\phi_i}$ and send all $\phi_i$ to zero, i.e.,
\begin{eqnarray}
    d\Delta_{N-1} \delta^4\left(P-\sum_{i=1}^Np_i\right)=\frac{d^N u}{U(1)^N}\delta(1-|\vec{u}|^2) &&= \frac{\prod_{i=1}^N r_idr_id\phi_i}{U(1)^N}\delta\left(1-\sum^N_{i=1}r_i^2\right)\nonumber \\
    &&= \int \prod_{i=1}^N\left[r_idr_id\phi_i \delta(\phi_i)\right]\delta\left(1-\sum^N_{i=1}r_i^2\right)\nonumber\\
    && = \frac{1}{2^N}\prod_{i=1}^N[d\rho_i]\left(1-\sum^N_{i=1}\rho_i\right),
\end{eqnarray}
where $\rho_i=r_i^2$. In~\cite{Larkoski:2020thc}, this is called the integration measure on the $(N-1)$-simplex. 

Furthermore, one can use $\delta^{2}(\vec{u}^\dagger \vec{v})$ to remove the integral of the last component of $v$, i.e.,
\begin{eqnarray}
    dS^{2N-3}=d^Nv \delta(1-|\vec{v}|^2)\delta^{2}(\vec{u}^\dagger \vec{v})=\frac{d^{N-1}v}{|u_N|^2}\ \delta\left(1-\sum_{i=1}^{N-1}|v_i|^2-|v_N|^2\right)\ .
\end{eqnarray}
From the condition $\vec{u}^\dagger \vec{v}=0$, we see that $v_N$ is a function depending on $v_{1,\dots, N-1}$ and $u_{1,\dots,N}$, i.e.,
\begin{eqnarray}
    v_N=-\frac{\sum_{i=1}^{N-1}u_i^* v_i}{u^*_{N}}=-\frac{\sum_{i=1}^{N-1}r_i v_i}{r_{N}}.
    \label{eq:v_to_vp}
\end{eqnarray}
To perform the integral over $v_{1,\dots, N-1}$, we need to make a symmetric real transformation for the variables $ v'=O^{-1} v$ such that $1-\sum_{i=1}^{N-1}|v_i|^2-|v_N|^2=1-\sum_{i=1}^{N-1}|v'|^2$, where the coordinates of $v'_{1,2,\cdots,N-1}$ are parametrized by
\begin{eqnarray}
    v_1'&=&e^{-i\xi_1}\cos\eta_1\\
    v_2'&=&e^{-i\xi_2}\sin\eta_1\cos\eta_2\nonumber\\
    \vdots\nonumber\\
    v_{N-2}'&=&e^{-i\xi_{N-2}}\sin\eta_1\dots\sin\eta_{N-3}\cos\eta_{N-2}\nonumber\\
    v_{N-1}'&=&e^{-i\xi_{N-1}}\sin\eta_1\dots\sin\eta_{N-3}\sin\eta_{N-2}.\nonumber
    \label{eq:}
\end{eqnarray}
Here $\xi_i\in [0,2\pi]$, $\eta_i\in [0,\pi/2]$. These coordinates can be viewed as embedding $S^{2N-3}$ into $\mathbb{C}^{N-1}$~\cite{Larkoski:2020thc}. 
Using these new variables, the relevant integral measure for $dS^{2N-3}$ is given by~\footnote{We identified a typo in the measure of $S^{2N-3}$ in Ref.~\cite{Larkoski:2020thc} and corrected them in eq.~\eqref{eq:veS}.}
\begin{eqnarray}
d^{N-1}v'\ \delta\left(1-\sum_{i=1}^{N-1}|v_i'|^2\right)=\left(\prod_{k=1}^{N-2}\cos\eta_k\sin^{2(N-2-k)+1}\eta_k\right)d\xi_i\dots d\xi_{N-1}d\eta_i\dots d\eta_{N-2}. \nonumber \\
\label{eq:veS}
\end{eqnarray}
After including the Jacobian for $v'=O^{-1} v$, we finally obtain the measure of $S^{2N-3}$.

From eq.~\eqref{eq:v_to_vp} and the condition $\sum_{i=1}^{N-1}|v_i|^2+|v_N|^2=\sum_{i=1}^{N-1}|v'|^2$, we find
\begin{eqnarray}
    (O^{-1})^2 = \mathbbm{1}+\frac{1}{r_N^2}\mathbf{r}^T\mathbf{r},
\end{eqnarray}
where $\mathbf{r}=(r_1,r_2,\dots,r_{N-1})$. 
The right-hand side of the equation takes the form of a rank-one modification of the identity matrix, whose inverse and square root can be computed analytically using the Sherman–Morrison Formula~\cite{zbMATH03055938}. For instance, 
\begin{eqnarray}
    &&\sqrt{\mathbbm{1}+\mathbf{u}^T\mathbf{u}} = \mathbbm{1}+\left(\frac{\sqrt{1+|\mathbf{u}|^2}-1}{|\mathbf{u}|^2}\right)\mathbf{u}^T\mathbf{u},\\
    &&\left(\mathbbm{1}+\mathbf{u}^T\mathbf{u}\right)^{-1} = \mathbbm{1}-\frac{1}{1+|\mathbf{u}|^2}\mathbf{u}^T\mathbf{u},
\end{eqnarray}
where $|\mathbf{u}|=\sqrt{\sum_i u_i^2}$ is the norm of the vector. 
Therefore, the matrix $O$ can be solved as
\begin{eqnarray}
    O=\mathbbm{1}-\left(\frac{\sqrt{1+\beta}-1}{\beta\sqrt{1+\beta}}\right)\frac{\mathbf{r}^T\mathbf{r}}{r_N^2},\ \ \beta=\frac{|\mathbf{r}|^2}{r_N^2}.
    \label{eq:invO}
\end{eqnarray}

\bibliographystyle{JHEP}
\bibliography{unitarity}

\end{document}